\documentstyle[emulateapj,graphicx]{article}

\begin{document}

\title{The radio afterglow from the giant flare of SGR 1900+14: the same mechanism
as afterglows from classic gamma-ray bursts? }

\author {K. S. Cheng$^1$  and X. Y. Wang$^2$}
\affil{$^1$Department of Physics, The University of Hong Kong, Hong Kong, China\\
$^2$ Department of Astronomy, Nanjing University, Nanjing 210093,
China\\}

\begin{abstract}
A radio afterglow was detected following the 1998 August 27 giant
flare from the soft gamma repeater (SGR) 1900+14. Its short-lived
behavior is quite different from the radio nebula of SGR 1806-20,
but very similar to radio afterglows from classic gamma-ray bursts
(GRBs). Motivated by this, we attempt to explain it with the
external shock model as invoked in the standard theory of GRB
afterglows. We find that the light curve of this radio afterglow
is not consistent with the forward shock emission of an
ultra-relativistic outflow, which is suggested to be responsible
for the initial hard spike of the giant flare. Nevertheless, shock
emission from a mildly or sub-relativistic outflow expanding into
the interstellar medium could fit the observations. The possible
origin for this kind of  outflow is discussed, based on the
magnetar model for SGRs. Furthermore, we suggest that the presence
of an ultra-relativistic fireball from SGR giant flares could be
tested by rapid radio to optical follow-up observations in the
future.
 \end{abstract}


\section{Introduction}
Soft gamma repeaters (SGRs) are generally characterized by
sporadic and short ($\sim0.1 {\rm s}$) bursts of hard X-rays with
luminosities as high as $10^4$ Eddington luminosity. They are also
well-know for two giant flares: the first on 1979 March 5 from SGR
0526-66 (Mazets et al. 1979) and the second on 1998 August 27 from
SGR 1900+14 (Hurley et al. 1999).  Frail et al. (1999) reported,
following the giant August 1998 flare from SGR 1900+14, the
detection of a transient radio source. Their observations covered
the time from about one week to one month after the flare. The
spectrum between 1 and 10 GHz is well fitted by a power-law with
$F_\nu\propto \nu^{-0.74\pm0.15}$. The source appears to have
peaked at about a week after the burst and subsequently undergone
a power law decay with an exponent of $\alpha=-2.6\pm1.5$.

The initial hard spike of the August 27 flare has a duration of
$\sim0.5 {\rm s}$ and luminosity greater than $2\times10^{44}{\rm
erg s^{-1}}$ ($>$15KeV) if the source distance is $d\simeq7 {\rm
Kpc}$ (Vasisht et al. 1994) . The short duration, high luminosity
and hard spectrum indicate that a relativistically expanding
fireball was driven from the star. The fireball should be
relatively clean and the Lorentz factor $\Gamma\ga10$ was inferred
from the luminosity and the temporal structure (Thompson \& Duncan
2001). With the experience of GRB afterglows, one may naturally
ask whether this power-law fading radio afterglow is due to the
blast wave emission driven by the fireball. Huang, Dai \& Lu
(1998) and Eichler (2003) had made some discussions on the
possible afterglow emission from SGRs.  In this paper, we try to
explain the radio afterglow from this flare. We study the
afterglow emission from the giant flare in section 2. We find that
shock emission from an ultra-relativistic outflow fails to explain
this radio afterglow. However, we propose that a mildly or
sub-relativistic outflow expanding into the interstellar medium
could fit the observations. Finally, we discuss the possible
origin for this kind of outflow in section 3.

\section{ Radio afterglow from SGR giant flares  }

We consider that an  outflow with ``isotropic" kinetic energy
$E_0$ and Lorentz factor $\Gamma_0$ ejected from the SGR  expands
into the ambient medium with a constant number density $n$. The
interaction between the  outflow and the surrounding medium is
analogous to GRB external shock (Rees \& M\'{e}sz\'{a}ros 1992;
M\'{e}sz\'{a}ros \& Rees 1997), but with quite different $E_0$ and
$\Gamma_0$. The Sedov time at which the shock enters the
non-relativistic phase is roughly given by
$t_{\rm nr}=\left({3E_0}/{4\pi n m_p c^5}\right)^{1/3}=1{\rm
days}\left({E_{0,44}}/{n_0}\right)^{1/3}$,
where $m_p$ is the proton mass and we used the usual notation
$a\equiv10^{n}a_n$. As the shock must have entered the
non-relativistic phase during the observation time of the radio
afterglow from the giant flare, we develop a model which holds for
both the relativistic and non-relativistic phases. From the view
of the energy conservation, the dynamic equation can be
approximately simplified as (e.g. Huang, Dai \& Lu 1999; Wang, Dai
\& Lu 2003)
\begin{equation}
(\Gamma-1)M_0 c^2 +  (\Gamma^2-1)m_{\rm sw} c^2=E_0,
\end{equation}
where $\Gamma$ is the Lorentz factors of the outflow, $m_{\rm
sw}=(4/3)\pi R^3 m_p n $ is the mass of the swept-up ISM ( $R$ is
the shock radius) and $M_0$ is the mass of the original outflow.
The first term on the left of the equation is the kinetic energy
of the outflow and the second term is the internal energy of the
shock.

The kinematic equation of the ejecta is
\begin{equation}
{dR}/{dt}={\beta c}/({1-\beta}),
\end{equation}
where $v=\beta c$  is the bulk velocity of the outflow with
$\beta(\Gamma)=(1-\Gamma^{-2})^{1/2}$ and $t$ is the observer
time. If the outflow is beamed and sideways expansion with sound
speed takes place, the expression of $m_{\rm sw}$ and the half
opening angle of the beamed outflow $\theta$ are respectively
given by
\begin{equation}
\frac{dm_{\rm sw}}{dt}=2\pi R^2(1-{\rm
cos}{\theta})\frac{nm_p\beta c}{1-\beta};~
\frac{d\theta}{dt}=\frac{c_s(\gamma+\sqrt{\gamma^2-1})}{R}
\end{equation}
where $c_s$ is the sound speed and we use the approximate
expression derived by Huang, Dai \& Lu (2000), which holds for
both the ultra-relativistic and non-relativistic limits.

Assuming that the distribution of the shock-accelerated electrons
takes a power-law form with the number density given by
$n(\gamma_e)d\gamma_e=K \gamma_e^{-p}d\gamma_e$ for
$\gamma_m<\gamma_e<\gamma_M$, the volume emissivity at the
frequency $\nu'$ in the comoving frame of the shocked gas is
(Rybicki \& Lightman 1979)
\begin{equation}
j_{\nu'}=\frac{\sqrt{3}q^3}{2 m_e c^2}\left(\frac{4\pi m_e c
\nu'}{3q}\right)^{\frac{1-p}{2}}B_\bot^{\frac{P+1}{2}}K
F_1(\nu',\nu'_m,\nu'_M),
\end{equation}
where $q$ and $m_e$ are respectively the charge and mass of the
electron, $B_\bot$ is the strength of the component of magnetic
field perpendicular to the electron velocity, $\nu'_m$ and
$\nu'_M$ are the characteristic frequencies for electrons with
$\gamma_m$ and $\gamma_M$ respectively, and
\begin{equation}
F_1(\nu',\nu'_m,\nu'_M)=\int^{\nu'/\nu'_m}_{\nu'/\nu'_M}F(x)x^{(p-3)/2}dx
\end{equation}
with $F(x)=x\int^{+\infty}_x K_{5/3}(t)dt$ ($K_{5/3}(t)$ is the
Bessel function). The physical quantities in the pre-shock  and
post-shock ISM are connected by the jump conditions (Blandford \&
Mckee 1976):$ n'=\frac{\hat{\gamma}\Gamma+1}{\hat\gamma-1}n$, $
e'=\frac{\hat{\gamma}\Gamma+1} {\hat{\gamma}-1}(\Gamma-1)nm_p{c^2}
$, where $e'$ and $n'$ are the energy and the number densities of
the shocked gas in its comoving frame and $\hat{\gamma}$ is the
adiabatic index, a simple interpolation of which between
ultra-relativistic and non-relativistic limits,
$\hat{\gamma}={(4\Gamma+1)}/{3\Gamma}$, gives a valid
approximation for trans-relativistic shocks.

Assuming that shocked electrons and the magnetic field acquire
constant fractions ($\epsilon_e$ and $\epsilon_B$) of the total
shock energy, we get
$\gamma_m=\epsilon_e\frac{p-2}{p-1}\frac{m_p}{m_e}(\Gamma-1)$,
$B_\bot=\sqrt{8\pi \epsilon_B e'}$ and $K=(p-1)n'\gamma_m^{p-1} $
for $p>2$. From the spectrum $F_\nu\propto \nu^{-0.74\pm0.14}$
 of the radio afterglow, we infer that $p\simeq2.5$. It is
reasonable to believe that $\nu'_M$, in comparison with the radio
frequencies, is very large throughout the observations. The
observer frequency $\nu$ relates to the frequency $\nu'$ in the
comoving frame by $\nu=D\nu'$, where $D={1}/{\Gamma(1-\beta {\rm
cos}\theta)}$ is the Doppler factor. The observed flux density at
$\nu$ is given by
\begin{equation}
F_\nu={V_{\rm eff} D^3 j_{\nu'}}/{4\pi d^2}
\end{equation}
where $V_{\rm eff}$ is the effective volume of the post-shock ISM
from which the radiation is received by the observer and should be
$V=m_{\rm sw}/n'm_p \Gamma^2$ for the isotropic case.

Below we shall study the afterglows from the relativistic outflow
with $\Gamma_0\sim10$ and the mildly-relativistic one
respectively.

\subsection{ Ultra-relativistic outflow }
{ In terms of Eqs(1)-(3), we can obtain the dynamic evolution of
the outflow, i.e. we get $\Gamma(t)$, $R(t)$, $m_{sw}(t)$ and
$\theta(t)$. Then using Eq.(4) and the expressions for $B_\bot$,
$K$ and $\gamma_m$, we can get the evolution of the observed flux
with time.} The radio afterglow of August 1998 flare peaks at
about one week after the burst. In the relativistic shock model,
it requires that the peak frequency $\nu_m$ crosses the
observation band at the peak time if the synchrotron radiation is
optically thin {\footnote {Numerical calculation with the formula
about the synchrotron self-absorption process in Wang, Dai \& Lu
(2000) shows that the synchrotron self-absorption optical depth at
the peak time is several orders of magnitude lower than unity. So,
crossing of the self-absorption frequency through the observation
band is not a viable explanation for this peak. }}. However, this
can hardly be satisfied for a ultra-relativistic outflow with
$\Gamma_0\sim10$, for reasonable values of the shock parameters
and the medium density $n$. Instead, the model light curves
generally peak at $t<0.1{\rm days}$. This can be clearly seen in
Fig. 1, in which we plot the model light curves with different
values of $n$ and $\epsilon_B$ for both the isotropic and beamed
outflow cases and compare them with the observation data.
Moreover, the peak flux $F_{\nu_m}$ are generally much larger than
the observed peak flux. The reason can be easily understood from
the following analytic estimate for the isotropic outflow case.

At the peak time  $t\sim 10 {\rm days}$ of the radio afterglow,
the shock had entered the non-relativistic phase, so the radius of
the shock is roughly $R\simeq (5/2)\beta c t$. From the condition
of energy conservation $\frac{4}{3}\pi R^3 \frac{\beta^2}{2}nm_p
c^2=E$, one can get $\beta=\left({12E}/{125\pi c^5 t^3 n
m_p}\right)^{1/5}=0.16E_{44}^{1/5}t_1^{-3/5}n_0^{-1/5}$.
 The magnetic field is
$B=\sqrt{8\pi\epsilon_B e'}=1.4\times10^{-3}{\rm G}
\epsilon_{B,-3}^{1/2}E_{44}^{1/5}t_1^{-3/5}n_0^{3/10}$ and the
peak frequency  and the peak flux are, respectively,given by
\begin{equation}
\nu_m=3\times10^5 {\rm Hz} (\frac{\epsilon_e}{0.3})^2E_{44}^{3/5}
t_1^{-9/5}n_0^{-1/10}\epsilon_{B,-3}^{1/2};
\end{equation}
\begin{equation}
F_{\nu_m}=\frac{N_e P_{\nu,m}}{4\pi d^2}=4.4\times10^7{\mu
Jy}E_{44}^{4/5}t_1^{3/5}n_0^{7/10}\epsilon_{B,-3}^{1/2}
\end{equation}
where $N_e$ is the total number of the swept-up electrons and
$P_{\nu,m}$ is the peak spectral power (Sari, Piran \& Narayan
1998). It is clearly seen that $\nu_m$ can hardly be as large as
$\nu_{\rm obs}=8.46{\rm GHz}$ for reasonable shock parameters of
$\epsilon_e$ and $\epsilon_B$ (e.g. Granot, Piran \& Sari 1999;
Wijers \& Galama 1999; Panaitescu \& Kumar 2002) and, furthermore,
the peak flux is much larger than detected from the giant flare.
Though this analytic estimate is for an isotropic outflow case,
the beamed outflow has also this problem as shown in Fig. 1.

Although the ultra-relativistic shock associated with the initial
hard spike of the giant flare could not be responsible for the
observed radio afterglow, we know from Fig. 1 that its radio
afterglow emission should be easily detected at the early time
even for the beamed case. The optical afterglow emission from the
ultra-relativistic shock is also calculated and shown in Fig.2.
Clearly, early optical afterglow emission can be as bright as
$100\mu {\rm Jy}$ { (R band magnitude $m_R=19$)} at $t\la0.1{\rm
days}$ for $\epsilon_B\ga10^{-3}$, a reasonable value we know from
GRB afterglows. So we urge early follow-up radio-to-optical
observations for future SGR giant flares to test the presence of
ultra-relativistic outflows. { Even at late time $t\sim10$ days,
the flux at  low radio frequencies are intense enough to be
detected. At, say,$\nu=150{\rm MHz}$, the extrapolated flux is
$0.4{\rm Jy} $ from Eqs(7) and (8)  for the typical parameters
used.}{\footnote { We estimate that the synchrotron
self-absorption frequency is below $10^7$Hz at this time.}} The
X-ray afterglow emission from the ultra-relativistic shock is,
however, predicted to be lower than the detected bright and pulsed
X-ray afterglow flux (Feroci et al. 2001), which is attributed to
the emission from the neutron star surface immediately after the
giant flare.

\subsection{Mildly or sub-relativistic outflow}
In \S2.1, we showed that the radio afterglow from the 1998 August
giant flare of SGR1900+14 is inconsistent with the shock emission
from an ultra-relativistic outflow. But a mildly or
sub-relativistic forward shock is able to explain the
observations, as we show below. The reason is that for a
mildly-relativistic or sub-relativistic outflow, it has an enough
long period of coasting phase of the outflow, during which the
flux increases with time even though $\nu_m\ll\nu_{\rm obs}$.
After this coasting phase, the shock decelerates and the flux
begins to fade in a power law manner. For an isotropic outflow,
the coasting phase lasts about
\begin{equation}
t\simeq\left(\frac{3E_0}{2\pi nm_p \beta_0^5 c^5}\right)^{1/3}
=5\times10^5 {\rm s}\,
E_{0,44}^{1/3}n_0^{-1/3}(\frac{\beta_0}{0.4})^{-5/3}
\end{equation}
until the mass of the ISM swept up by the blast wave is comparable
to the mass of the outflow, where $\beta_0$ is the initial
velocity of the outflow. During the coast phase, $\beta=\rm const$
and $R=\beta t\propto t^1$. According to Eqs.(4) and (6),
$F_\nu\propto R^3 \beta^{(5p-3)/2}\propto t^3$. When the mildly or
sub-relativistic outflow is decelerated by the swept-up mass, it
quickly enters the Sedov phase, during which
$\beta\propto{t^{-3/5}}$ and $R\propto t^{2/5}$. So, $F_\nu\propto
t^{(21-15p)/10}\propto t^{-1.65}$ for $p=2.5$. For a beamed
outflow, the coasting phase is similar to the isotropic case, as
the expansion is dominated by the cold ejecta during this phase.
But during the deceleration phase, the shocked ISM plasma has
comparable energy to initial energy $E_0$ and the sideways
expansion may take place, so it is expected that the flux may
decay steeper than the isotropic case (e.g. Rhoads 1999; Sari,
Piran \& Halpern 1999).

The fits with model light curves for mildly or sub-relativistic
outflow are presented in Figs. 3 and 4 for different shock
parameters. In Fig.3, the isotropic energy is chosen to be
$E_{0,\rm iso}=10^{44}{\rm erg}$ while in Fig 4 the real energy of
the ejecta is chosen to be $10^{44}{\rm erg}$ (its isotropic
energy is therefore $10^{44}/(\theta_j^2/2){\rm erg}$,
where$\theta_j$ is the initial half opening angle). The beamed
outflow model can provide nice fits of the observations for a wide
range of shock parameters such as $n$ and $E_0$. In Fig.3, we also
present the model light curve for an isotropic outflow (without
sideways expansion) denoted with the dashed line. Clearly, it
decays too slowly to fit the observations. In all these fits, we
used fixed values for $E$, $p$, $\theta_j$, $\epsilon_e$ and $n$
with only two free parameters: the initial Lorentz factor
$\Gamma_0$ and $\epsilon_B$. We therefore conclude that the mildly
or sub-relativistic outflow from the SGR giant flare could provide
a plausible explanation for this radio afterglow.

\section{Discussions and Conclusions}
We have shown that a mildly or sub-relativistic outflow from the
SGR could be consistent with this radio afterglow. This outflow is
expected to originate from the neutron star curst, accompanying
the giant flare. SGRs are now believed to be "magnetars", neutron
stars with surface field of order $10^{14}-10^{15}$ Gauss or more
(Duncan \& Thompson 1992;  Thompson \& Duncan 1995). The energy
that drives the very large ($\ga10^{44}{\rm erg}$) giant flares
such as the March 5, 1979 event from SGR 0526-66 and the August
27, 1998 event from SGR 1900+14 is attributed to a sudden
large-scale rearrangement of the magnetic field which releases the
magnetic energy, while the smaller short repeating bursts
($E\la10^{41}{\rm erg}$) seem to be well explained as being driven
by the localized yield of the crust to the magnetic strain. A
magnetic field with $B>(4\pi \phi_{\rm max} \mu)^{1/2}\sim
2\times10^{14}(\phi_{\rm max}/10^{-3})^{1/2}{\rm G}$ can fracture
the curst, where $\mu\sim 10^{31}(\rho/\rho_{\rm nuc})^{4/3}{\rm
erg cm^{-3}}$ is the shear modulus of the curst and $\rho_{\rm
nuc}$ is the nuclear density and $\phi_{\rm max}$ is the yield
strain of the crust. However, such a patch of crust
 is too heavy to be able to
overcome the binding energy of the neutron star. We expect that
only a tiny fraction of the fracturing curst matter can overcome
the gravitational binding energy and is able to be accelerated to
a mildly-relativistic velocity $\Gamma_0-1\sim0.1$ by the released
magnetic field energy. Note that the kinetic energy of the
mildly-relativistic matter per unit of mass $(\Gamma_0-1)c^2$ is
comparable to the binding energy $GM_{\rm NS}/R$, where $M_{\rm
NS}$ and $R$ are the mass and radius of the neutron star
respectively. Let's denote the amount of matter as $\Delta m$, the
isotropic kinetic energy $E_0$ and the real energy of the beamed
outflow $E_{\rm r}=E_0\theta_j^2/2$, where $\theta_j$ is the
beaming angle of this outflow. For $\Gamma_0-1\sim0.1$, $\Delta
m=5\times10^{22}E_{{\rm r},43}\,{\rm
g}=5\times10^{23}E_{0,44}\theta_j^2\,{\rm g}$.  Let the size of
this patch of matter be $\Delta r$. Because of the insensitivity
of $\Delta r$ on $\Delta m$ for the outermost curst of neutron
star, we estimate $\Delta r\simeq0.1-0.3{\rm Km}$ for $E_{\rm
r}\simeq 10^{42}-10^{44}{\rm erg}$.

Once we know $\Delta r$, we can estimate the beaming angle
$\theta_j$ of the outflow when it breaks away from the confinement
of the magnetic field. This amount of matter will be vaporized and
become plasma near the neutron star surface, which moves out along
the open field lines of the magnetar. The initial kinetic energy
density  of this outflow is $\varepsilon_{k0}=\dot{E_{\rm r}}/(A_0
\beta_0 c)=1.6\times10^{24}{\rm erg cm^{-3}}\dot E_{{\rm r},43}
(\Delta r/2{\rm Km})^{-2}(\beta_0/0.4)^{-1}$, where $\dot E_{\rm
r}$ is the real kinetic energy luminosity of this outflow,
$\beta_0 c$ is the initial velocity of this outflow and $A_0$ is
the initial sectional area. As the plasma moves out to radial
radius $r$, the sectional area $A=\pi (r{\rm sin}\theta)^2$, where
$\theta$ is the angle relative to the magnetic axis. Because the
magnetic field lines satisfy $r\propto {\rm sin}^2\theta$,
$A\propto r^3$ for small $\theta$ and $\varepsilon_{k}\propto
r^{-3}$. On the other hand, the magnetic field energy density
scales with $r$ as $\varepsilon_{B}=(B_0^2/8\pi)(r/R)^{-6}$, where
$B_0$ is the surface magnetic field of the neutron star. When the
magnetic field energy density decreases to be comparable to the
kinetic energy density, the outflow plasma breaks from the
confinement of the magnetic field. This corresponds to a radius
$r_b/R\simeq (B_0^2/8\pi \varepsilon_{k0})^{1/3}\simeq 30
B_{0,15}^{2/3}\dot E_{{\rm r},43}^{-1/3}(\beta_0/0.4)^{1/3}(\Delta
r/0.2{\rm Km})^{2/3}$ and
$\theta_b/\theta_0\simeq(r/R)^{1/2}\simeq 5.5 B_{0,15}^{1/3}\dot
E_{{\rm r},43}^{-1/6}(\beta_0/0.4)^{1/6}(\Delta r/0.2{\rm
Km})^{1/3}$. Note that $\theta_b$, which is roughly equal to the
beaming angle $\theta_j$, is very insensitive to the value of
$\dot E_{{\rm r}}$. As the initial opening angle near the neutron
star surface $\theta_0=\Delta r/R=0.02(\Delta r/0.2 {\rm Km})$, we
estimate the beaming angle of the outflow is
$\theta_j\simeq0.1-0.2$ for typical parameters.

What powers the ejection of this patch of matter? We think that
the reconnection of the magnetic field within a  region  of size
$\Delta r$  during the period of the giant flare will release
energy of $(B^2/8\pi)(\Delta r)^2 V_{\rm A} \Delta
t\sim(B^2/8\pi)(\Delta r)^2 R$, which should be equal to $\Delta m
(\Gamma_0-1)c^2$, where $B$ is the crust magnetic field, $V_{\rm
A}$ is the internal Alfv\'{e}n velocity and $\Delta t$ is the
growth time of the instability causing the giant flare, viz. the
duration ($\Delta t\simeq0.5{\rm s}$) of the initial hard spike of
the August 27 giant flare(Thompson \& Duncan 1995, 2001). The size
is therefore estimated to be $\Delta r\sim 0.2{\rm Km}$ and the
mass is $\Delta m\sim10^{23}B_{15}^2\,{\rm g}$, which are in
reasonable agreement with the above estimates  according to the
light curve fits if $E_0\sim10^{45}{\rm erg}$ or equally $E_{\rm
r}\sim10^{43}{\rm erg}$.

{ Note that continuing acceleration of electrons at the surface of
the neutron star is also a possible mechanism for the radio
afterglow, and need further careful investigation in future.}

In summary, we studied the afterglow emission from the possible
ultra-relativistic outflow and mildly or sub-relativistic outflow
accompanying the SGR giant flares. The radio afterglow emission
from the August 27 giant flare of SGR 1900+14 is consistent with a
mildly or sub-relativistic outflow, but could not be produced by
the forward shock emission from an ultra-relativistic outflow.
However, we predict that this ultra-relativistic outflow,
suggested to be associated with the hard spike of the giant flare,
if exist, should produce bright radio to optical afterglows at the
early phase ($t\la0.1 {\rm days})$, which can be tested by future
observations.

{\acknowledgments We would like to thank the referee for valuable
comments and thank Z. G. Dai, Y. F. Huang and T. Lu for useful
discussions. This work was supported by a RGC grant of Hong Kong
government, the Special Funds for Major State Basic Research
Projects and the National Natural Science Foundation of China
under grants 10233010 and 10221001. }

\clearpage

\begin{figure*}[t]
\plotone{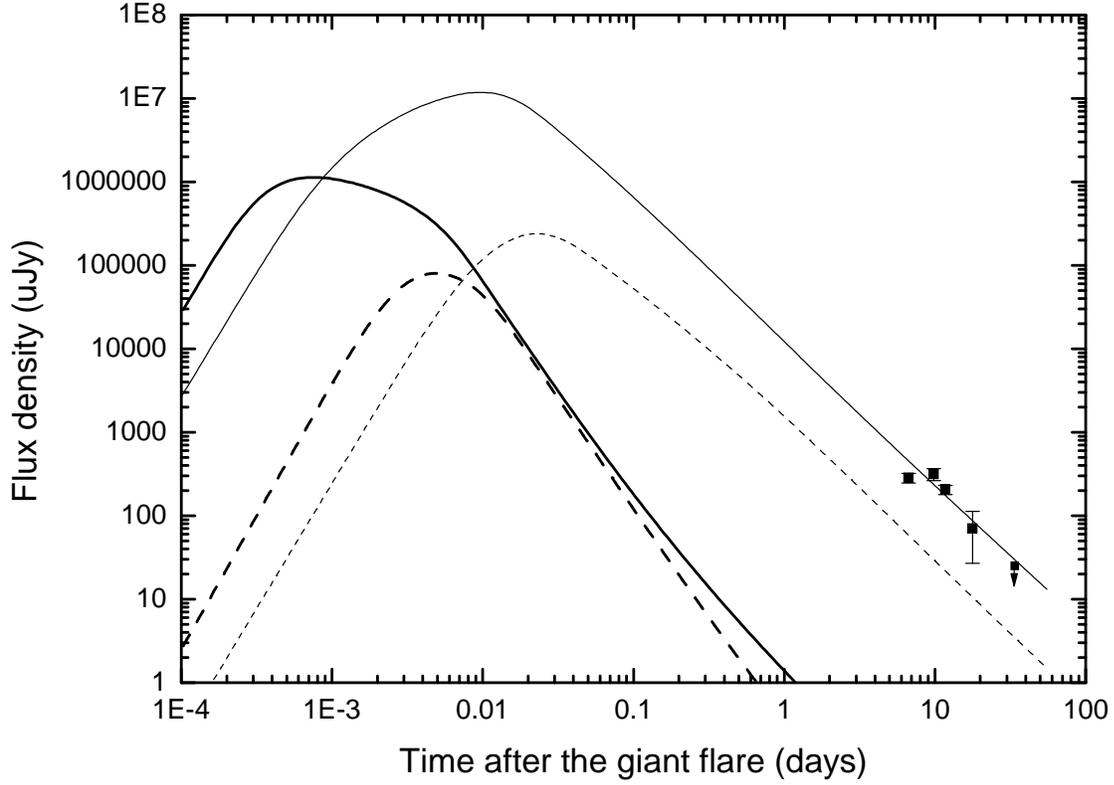}
\caption{Comparison between the model light curves of the
afterglows from ultra-relativistic outflows ($\Gamma_0=10$) with
the observations of the radio flare from the August 27 giant flare
of SGR 1900+14 { observed at the frequency 8.46GHz}. Detections
and upper limits for the non-detections, taken from Frail et al.
(1999), are indicated by the filled squares and arrows
respectively. The {\em thin} solid and dashed lines represent the
afterglows of isotropic outflows expanding into ISM with $n=1{\rm
cm^{-3}}$ and $n=0.01{\rm cm^{-3}}$ respectively. Other parameters
used are $E_{\rm iso}=10^{44}{\rm erg}$, $\epsilon_e=0.3$,
$\epsilon_B=10^{-5}$. The {\em thick} solid and dashed lines
represent the afterglows of beamed outflows with $\theta_j=0.15$
expanding into ISM with $n=1{\rm cm^{-3}}$ and $n=0.01{\rm
cm^{-3}}$ respectively. Other parameters are the same as thin
lines. }
\end{figure*}
\begin{figure*}[t]
\plotone{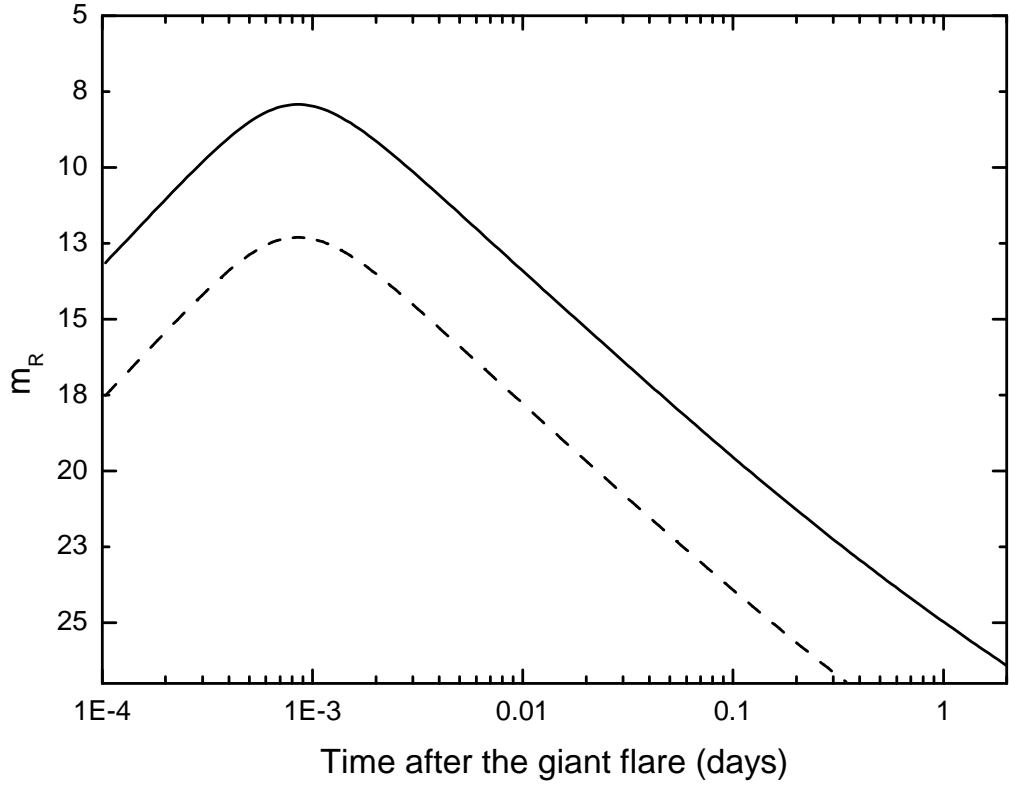}
\caption{Predicted $R$-band ($\nu_R=4.4\times10^{14}{\rm
Hz}$)optical afterglow light curves of beamed ($\theta_j=0.15$)
and ultra-relativistic ($\Gamma_0=10$) outflows from SGR giant
flares with $E_{\rm iso}=10^{44}{\rm erg}$, $n=1{\rm cm^{-3}}$ and
$\epsilon_e=0.3$, but with different values for $\epsilon_B$. The
solid and dashed lines correspond to $\epsilon_B=10^{-3}$ and
$\epsilon_B=10^{-5}$ respectively. }
\end{figure*}
\clearpage
\begin{figure*}
\plotone{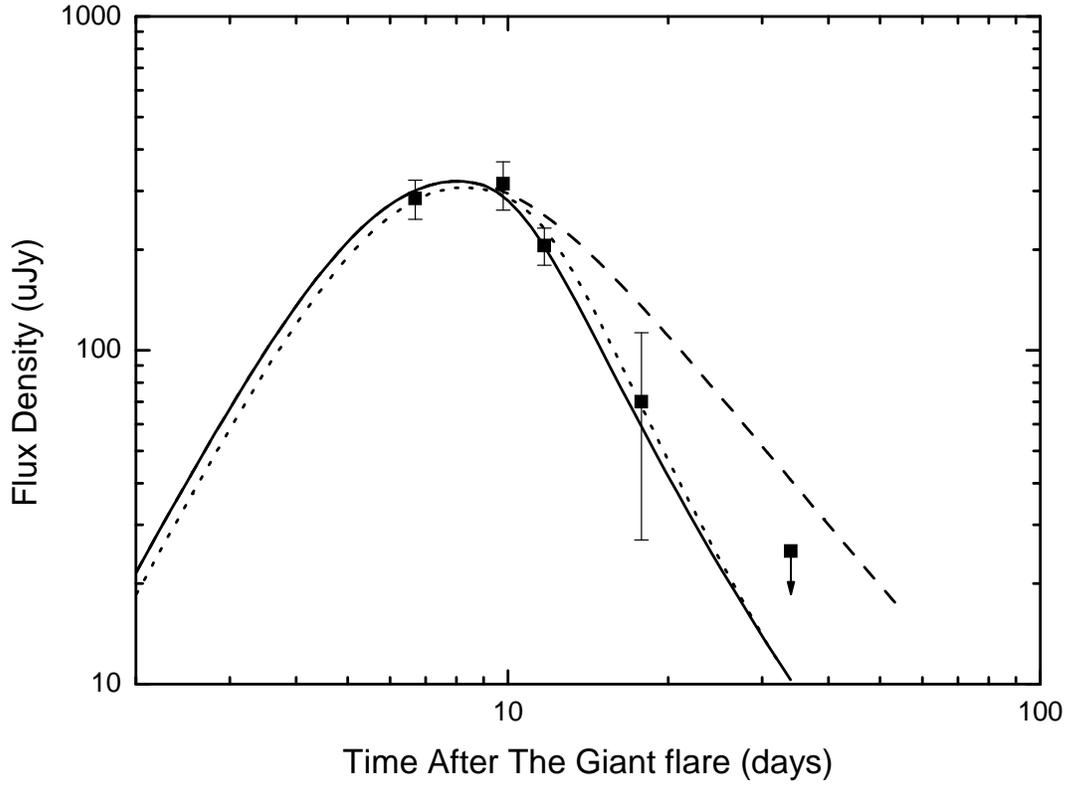}
\caption {Fits of the radio (8.46GHz) flare  from August 27 SGR
giant flare with afterglow emission from beamed, mildly or
sub-relativistic outflow with isotropic energy $E_{\rm
iso}=10^{44}{\rm erg}$ and $\theta_j=0.15$.  The solid and dotted
lines correspond to beamed outflows with sideways expansion
expanding into ISM with $n=1{\rm cm^{-3}}$ and $n=0.01{\rm
cm^{-3}}$ respectively. Other parameters used are, respectively,
($\Gamma_0=1.033$, $\epsilon_e=0.3$, $\epsilon_B=0.008$) and
($\Gamma_0=1.13$, $\epsilon_e=0.3$, $\epsilon_B=0.03$). The dashed
line have same parameters as the solid line except that no
sideways expansion is considered. }
\end{figure*}

\begin{figure*}[t]

\plotone{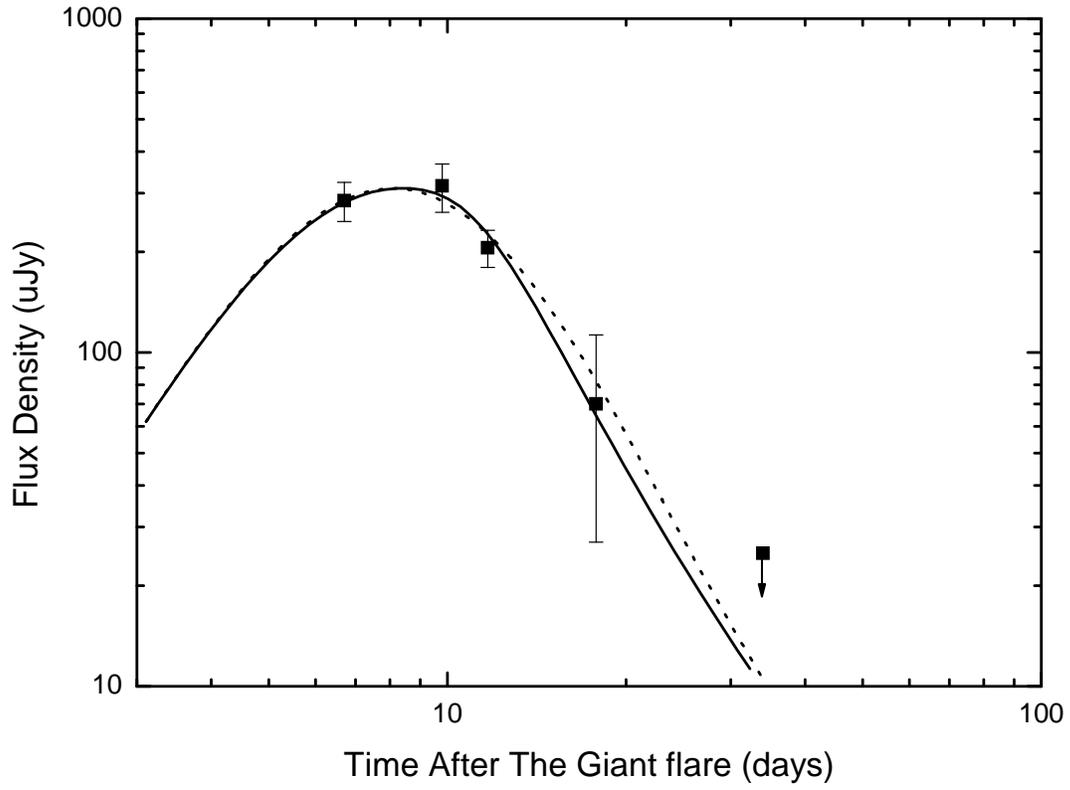}

\caption{Same as Fig. 3 but with $E_{\rm
iso}\theta_j^2/2=10^{44}{\rm erg}$. Parameters used are ($n=1{\rm
cm^{-3}}$, $\Gamma_0=1.13$, $\epsilon_e=0.3$,
$\epsilon_B=3\times10^{-6}$) for the solid line and ($n=0.01{\rm
cm^{-3}}$, $\Gamma_0=1.4$, $\epsilon_e=0.3$,
$\epsilon_B=1.5\times10^{-5}$) for the dotted line, respectively.}
\end{figure*}

\end{document}